\documentclass[aps,prc,nofootinbib,floatfix,superscriptaddress,longbibliography,reprint]{revtex4-2}
\usepackage[T1]{fontenc}
\usepackage[utf8]{inputenc}
\usepackage[english]{babel}

\usepackage{graphicx}
\usepackage{isotope}
\usepackage{physics}
\usepackage{xcolor}
\usepackage{hyperref}
\usepackage{amsmath}
\usepackage{multirow}
\usepackage{xspace}
\usepackage[nameinlink]{cleveref}

\usepackage{times}
\usepackage{txfonts}

\usepackage{siunitx}
\DeclareSIUnit{\fm}{\femto\meter}
\DeclareSIUnit{\electron}{\text{e}}
\DeclareSIUnit{\quadrupole}{\electron\fm\squared}
\DeclareSIUnit{\sqquadrupole}{\electron\squared\fm\tothe{4}}
\DeclareSIUnit[quantity-product = {}]{\percent}{\%}
\crefname{table}{Tab.}{Tabs.}
\Crefname{table}{Table}{Tables}
\crefname{figure}{Fig.}{Figs.}
\Crefname{figure}{Figure}{Figures}
\crefname{section}{Sec.}{Secs.}
\Crefname{section}{Section}{Sections}
\crefname{equation}{Eq.}{Eqs.}
\Crefname{equation}{Equation}{Equations}

\DeclareMathAlphabet{\curly}{OMS}{cmsy}{m}{n}

\newcommand*{\elem}[2]{\ensuremath{\isotope[#2]{\mathrm{#1}}}}

\newcommand*{\hw}{\ensuremath{{\hbar\Omega}}\xspace}
\newcommand*{\aHO}{\ensuremath{{a_\mathrm{HO}}}\xspace}
\newcommand*{\Nmax}{\ensuremath{{N_\mathrm{max}}}\xspace}
\newcommand*{\cNmax}{\ensuremath{{\curly{N}_\mathrm{max}}}\xspace}

\definecolor{FGOrange}{rgb}{1.0, 0.5490196078431373, 0.0}
\definecolor{FGgreen}{rgb}{.0, 0.5, 0.0}
\definecolor{FGred}{rgb}{1.0, 0.0, 0.0}

\definecolor{FGredgray1}{rgb}{.5, 0.5, 0.5}
\definecolor{FGredgray2}{rgb}{.6016, 0.3359, 0.3359}
\definecolor{FGredgray3}{rgb}{.7031, 0.1680, 0.1680}
\definecolor{FGredgray4}{rgb}{.6797, 0.0742, 0.0742}
\definecolor{FGredgray5}{rgb}{.5586, 0.0352, 0.0352}
\definecolor{FGredgray6}{rgb}{.4492, 0., 0.0}

\begin{document}

\title{Machine Learning for Correlations of Electromagnetic Properties in Ab Initio Calculations}

\author{Marco~Kn\"oll}
\email{marco.knoell@physik.tu-darmstadt.de}
\affiliation{Institut f\"ur Kernphysik, Fachbereich Physik, Technische Universit\"at Darmstadt, Schlossgartenstr. 2, 64289 Darmstadt, Germany}
\author{Marc~L.~Agel}
\affiliation{Institut f\"ur Kernphysik, Fachbereich Physik, Technische Universit\"at Darmstadt, Schlossgartenstr. 2, 64289 Darmstadt, Germany}
\author{Tobias~Wolfgruber}
\affiliation{Institut f\"ur Kernphysik, Fachbereich Physik, Technische Universit\"at Darmstadt, Schlossgartenstr. 2, 64289 Darmstadt, Germany}
\author{Pieter~Maris}
\affiliation{Dept.\ of Physics and Astronomy, Iowa State University, Ames, Iowa 50011, USA}
\author{Robert~Roth}
\email{robert.roth@physik.tu-darmstadt.de}
\affiliation{Institut f\"ur Kernphysik, Fachbereich Physik, Technische Universit\"at Darmstadt, Schlossgartenstr. 2, 64289 Darmstadt, Germany}
\affiliation{Helmholtz Forschungsakademie Hessen f\"ur FAIR, GSI Helmholtzzentrum, 64289 Darmstadt, Germany}

\date{\today}

\begin{abstract}

\noindent In ab initio nuclear structure theory, accurately predicting electromagnetic observables, such as moments and transition rates, is essential for a comprehensive understanding of nuclear properties.
However, computational limitations and conceptual difficulties often hinder the precise calculation of these observables.
In this work, we extend machine learning methods for model-space extrapolations to electric quadrupole moments. 
We further present a new machine learning approach that leverages the correlations between energies, radii, and electromagnetic observables.
By learning these correlations from no-core shell model calculations in accessible model spaces, this new model enables the prediction of electromagnetic observables in the full Hilbert space from predictions of full-space energies and radii, which can be obtained with established machine learning extrapolation tools.
An essential property of our approach is the capability for uncertainty quantification, allowing for reliable predictions with combined statistical error estimates for many-body and interaction uncertainties.
Being solely built upon the physical correlations of different observables, it can be generalized across different ab initio methods.
We demonstrate the power of this new extrapolation scheme through a precision study of electric quadrupole moments across a wide range of p-shell nuclei.

\end{abstract}

\maketitle

\section{Introduction}

Ab initio nuclear structure theory has entered a precision era with various many-body methods being able to describe energy spectra and radii with good precision across a wide range of nuclei.
However, the description of electromagnetic (EM) properties, such as the electric quadrupole (E2) moment, remains notoriously difficult either due to the non-scalar nature of the corresponding operators, which poses conceptual challenges for various medium-mass methods \cite{PaSt17,Vobig2020,StHe22,BoBa24,RoPe24}, or due to slow convergence with model-space size in basis expansion methods such as the no-core shell model (NCSM) \cite{BaNa13,NaQu09,Roth09,ZheBa93,FoRo13}.
The issue of incomplete convergence is not exclusive to EM observables and has been subject to a multitude of studies featuring different attempts to extrapolate calculations from accessible model spaces to the infinite Hilbert space \cite{Roth09,MaVa09,BoFu07b,FuHa12,CoAv12,MoEk13,FuMo14,WeFo15}.
More recently, this problem has been addressed with machine learning models, which have proven very successful for the extrapolation of ground- and excited-state energies as well as radii in the NCSM \cite{NeVa19,JiHa19,KnoWo23,WoKno24}.
In addition, they provide ways to extract meaningful uncertainties that are crucial for precision calculations of nuclear properties.

In this work, we extend the method presented in \cite{WoKno24} to the E2 moment.
We train artificial neural networks (ANNs) to capture the observable-specific convergence pattern from fully converged NCSM calculations of few-body systems, i.e.\ $A\leq4$.
Subsequently, these ANNs can be used to predict the converged value for heavier nuclei based on calculations in small model-spaces.
This essentially resembles a conversion of the extrapolation problem into an interpolation problem, which is key to the success of this approach.
As these ANNs predict the solution in the full Hilbert space, we refer to them as full-space prediction networks (FSPNs).
In principle, this method can be applied to every observable, however, most EM observables are non-existent in the few-body systems required for training.
In fact, for the E2 moment we are limited to training data for \elem{H}{2}.

In order to circumvent these limitations, we further propose a new ANN architecture that builds on correlations between observables.
Instead of learning the convergence pattern we use NCSM data to capture the dependence of EM observables on energy and radius, which is specific to nucleus, interaction and state.
In case of the E2 moment it has been shown that it strongly correlates with the radius, which is also implied by the analytical expressions of the respective operators \cite{CaFa22,CaMa24}.
Our new approach is designed to capture these correlations in an ANN that is then used to predict the full-space solution of the E2 moment from FSPN predictions of energy and radius.
We refer to those ANNs as observable transcoder networks (OTNs).
As these OTNs solely capture the correlations between the observables, independent of a nuclear many-body model, the full-space energy and radius required to predict the full-space E2 moment can be obtained by any means.
Hence, a generalization to other methods is naturally given.

After a detailed discussion of the model-space extrapolation we proceed to incorporate interaction uncertainties based on a Bayesian uncertainty procedure developed by the BUQEYE collaboration \cite{MeFu19} and perform a precision study of the E2 moments in selected p-shell nuclei.
Finally, we give a brief outlook on cross-method applications.

\section{Basics}

\subsection{No-Core Shell Model}

The NCSM gives access to a wide range of observables in both ground states and excited states.
We expand the eigenstates $\ket{\Psi_n}$ of the nuclear Hamiltonian $H$ in a basis of Slater determinants $\qty{\ket{\phi_i}}$, thus, transforming the nuclear many-body problem into a matrix eigenvalue problem
\begin{align}
    \sum_j\mel*{\phi_i}{H}{\phi_j}\braket*{\phi_j}{\Psi_n} = E_n\braket{\phi_i}{\Psi_n} \quad \forall i
\end{align}
with energy eigenvalues $E_n$.
The Slater determinants themselves are constructed from harmonic oscillator (HO) single-particle states.
This HO basis comes with a length scale \aHO that can be converted into the HO frequency \hw.
In order to render this infinite dimensional matrix eigenvalue problem finite, we truncate the basis by limiting the number of HO excitation quanta in a Slater determinant to \Nmax.
When starting in small model spaces and successively increasing \Nmax, a convergence pattern for the eigenvalues emerges.
This convergence pattern is determined by \Nmax and \hw.
In the limit $\Nmax\rightarrow\infty$ the exact solution in the full Hilbert space is recovered, independent of \hw.

For different observables the convergence pattern and the rate of convergence can be drastically different.
While energies show a monotonously decreasing behavior due to the variational character of the NCSM, radii or EM observables have no such constraints and exhibit ascending, descending, or even oscillating sequences.
This severely complicates the extrapolation to infinite model spaces. 

The convergence pattern is further influenced by the choice of the interaction.
In this work, we use realistic nucleon-nucleon (NN) and three-nucleon (3N) interactions from chiral effective field theory (EFT) \cite{MaEn11,EpHa09,HuVo20,LENPIC21,HeBo13}, which are grouped into families based on the regulator scheme.
Within each family, interactions are available at various chiral orders and for different cutoffs $\Lambda$.
Further, an SRG transformation of the Hamiltonian that accelerates the convergence rate is employed \cite{BoFu07,BoFu10,RoNe10,FuHe13,RoCa14}.
Note that we do not include SRG corrections for the radius and E2 operator as these are typically below \SI{1}{\percent} and, therefore, negligible compared to the many-body and interaction uncertainties.

Overall, there is a multitude of factors that impact the convergence pattern, which, on one hand, further complicates the extrapolation problem, but, on the other hand, allows us to generate a wealth of training data for machine-learning approaches.

\begin{figure}
    \centering
    \includegraphics[width=\columnwidth]{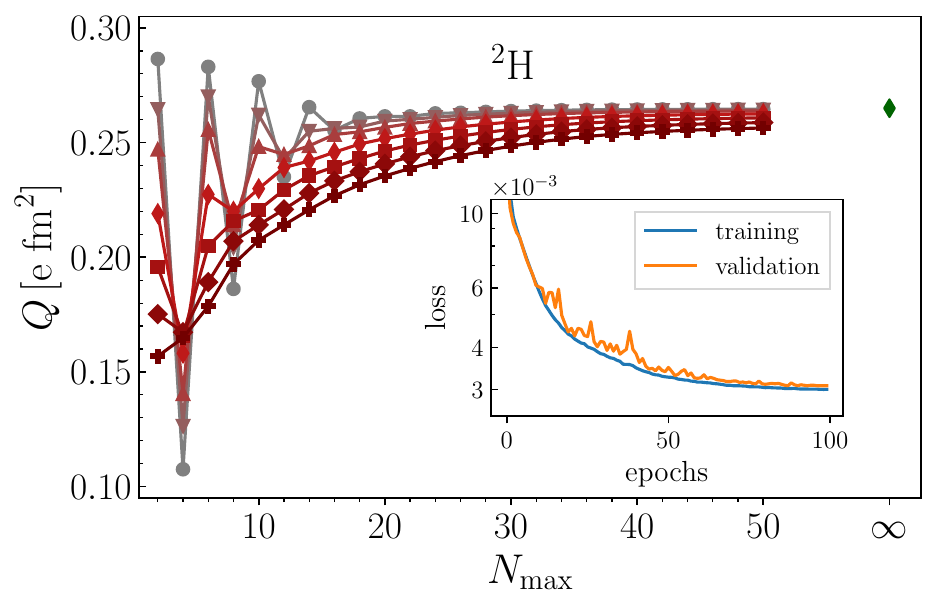}
    \caption{Example set of NCSM calculations for the E2 moment of \elem{H}{2} for the EMN[500] interaction with converged value obtained at $\Nmax=200$ used as FSPN training data. Inset: Loss on the training and validation data during the training of an FSPN averaged over 10 networks.}
    \label{fig:training_data}
\end{figure}

\subsection{Artificial Neural Networks}

In general, ANNs can be considered universal function approximators that excel at interpolation tasks.
Since the exact functional form of NCSM convergence patterns is unknown, they provide the ideal tool to model these patterns.

ANNs are a class of machine learning models inspired by the structure and functioning of biological neural networks. 
At their core, ANNs consist of layers of interconnected nodes, or neurons, that process input data and learn to map complex relationships between inputs  and outputs \cite{goodfellow2016deep}. 
These neurons are organized into an input layer, one or more hidden layers, and an output layer. 
Each neuron applies a weighted sum of its inputs $x^\mathrm{in}$ and adds a bias, followed by a non-linear activation function~$\sigma$
\begin{flalign}
    x^\mathrm{out}_i=\sigma\qty(\sum_j x^\mathrm{in}_j w_{ji} + b_i),
\end{flalign}
which allows the network to model highly non-linear relationships in the data.
Here, $w_{ij}$ and $b_i$ denote the free parameters of the ANN model.

Training an ANN involves adjusting these parameters based on a given set of training data. 
This is typically done using a process called back-propagation \cite{Rojas96}, in which the deviation of the network’s predictions from the true outputs, also called loss, is propagated backward through the network. 
The parameters are then updated iteratively using optimization techniques such as stochastic gradient descent to minimize the loss, effectively enabling the network to learn from the data.

We can leverage these adaptive capabilities and use ANNs to model NCSM convergence patterns or capture correlations between observables.
Note that the actual implementation of an ANN and its training requires various choices, most importantly, network size, activation function, loss function, optimization algorithm, and learning rate.
These so-called hyperparameters greatly influence the performance of the ANNs and we will discuss them in more detail for FSPNs and OTNs, respectively.

\section{Full-Space Prediction Networks}\label{sec:FSPN}

We start our investigations of E2 moments by applying the FSPN approach presented in \cite{WoKno24}.
Conceptually, we attempt to capture the convergence pattern in few-body ($A\leq 4$) calculations, which can be fully converged in the NCSM.
Given the assumption that these patterns are universal across different nuclei, we can employ the FSPN to predict the converged results for a wide range of p-shell nuclei from non-converged calculations at small \Nmax.
Previous applications to energies and radii have demonstrated that this assumption generally holds true.
For the training we typically use data for \elem{H}{2}, \elem{H}{3}, and \elem{He}{4}.
However, only \elem{H}{2} exhibits a non-vanishing E2 moment.
Thus, in contrast to previous applications on energies and radii \cite{KnoWo23,WoKno24}, the training data is restricted to a single nucleus.
This severely limits the diversity of convergence patterns in the training data.
Nevertheless, we can exploit different Hamiltonians, SRG flow parameters, and HO frequencies to generate a variety of data.
Our training data consists of NCSM results for \elem{H}{2} up to $\Nmax=50$ for $\hw=\qtylist{12;14;16;20;24;28;32}{\MeV}$ calculated with the non-local chiral NN interactions from \cite{EnMa17,HuVo20} at  N$^2$LO, N$^3$LO, and N$^4$LO' for cutoffs $\Lambda = \qtylist{450;500;550}{\MeV}$.
In addition to the bare interactions, we employ SRG evolutions up to $\alpha=\qtylist{0.02;0.04;0.08}{\fm\tothe{4}}$.

\Cref{fig:training_data} shows a typical example for such a set of training data for a single Hamiltonian. 
The target values for the training are obtained at $\Nmax = 200$ at which the calculations are sufficiently converged.
A striking feature is the strongly oscillating behavior in model-spaces up to $\Nmax\approx18$, which can also occur for p-shell nuclei but is particularly pronounced in \elem{H}{2}.  
This demonstrates the complicated character of the convergence patterns for E2 moments and the difficulties for the extrapolation scheme.
We emphasize that in the selected frequency range for the training data, the overall direction of convergence is ascending towards larger values. 
We will address this again in the context of negative E2 moments.

\subsection{Network Design}

\begin{figure}
    \centering
    \includegraphics[width=.8\columnwidth]{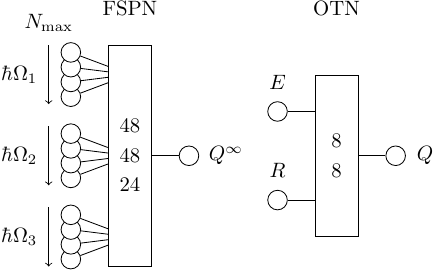}
    \caption{Schematic topologies of an FSPN and an OTN. The numbers in the boxes indicate number and size of the hidden layers.}
    \label{fig:topology}
\end{figure}

The network design in this work is identical to previous FSPN applications.
We predict the solution in the full Hilbert space based on a subset or sample $S^{\cNmax}$ of the available NCSM data consisting of E2 moments at four consecutive \Nmax for three different \hw while \cNmax indicates the highest \Nmax present in the sample. 
The desired mapping $S^\cNmax\rightarrow Q^\infty$ is then modeled by an ANN with an input layer, three hidden layers, and an output layer with dimensions (12, 48, 48, 24, 1), where the node in the output layer yields the prediction for the full-space E2 moment $Q^\infty$. 
A schematic depiction of the network topology is given in \cref{fig:topology} and contains 4189 trainable parameters.
Regarding the chosen hyperparameters we employ an AdamW optimizer \cite{Losh17}, a ReLU activation function \cite{Nair10}, and a mean-square error (MSE) loss function.

For the training of the network we construct all possible samples from the training data, which amounts to 166\,320 samples, and randomly separate it into a large training set (85\%), a small validation set (10\%), and a development set for adjustments during training (5\%).
In order to facilitate the learning process, we normalize each sample to the interval [0,1], which has proven to be crucial for the prediction of radii \cite{WoKno24}.
We then train our FSPN on batches of 512 training samples with an initial learning rate of $\eta=0.001$, which is successively reduced during the training by an adaptive learning rate scheduler that halves the learning rate whenever it plateaus for more than two epochs.
The network is trained for a total of 100 epochs and is considered valid if its loss on the validation data is below $0.01$.
The inset in \cref{fig:training_data} shows the loss of the FSPNs for each epoch on, both, the training and validation data averaged over 10 networks.
From the proximity of both curves we conclude that the FSPNs do not suffer from overfitting.

\subsection{Statistical Uncertainties}

For the evaluation of the FSPN we decompose the evaluation data analogously to the sample construction for the training data in order to match the input layer of the network.
All samples are passed through the trained network, each yielding a prediction of the full-space E2 moment.
This naturally results in a distribution of predictions conveniently represented by a histogram.
It turns out that position and width of the distribution change significantly if sequences for unnatural, in particular very small, values of \hw are included in the evaluation set.
In many cases this even results in multi-modal distributions as these sequences often exhibit strongly oscillating behavior or extremely slow convergence.
We, therefore, restrict ourselves to values of \hw that are close to the optimum and show regular, ideally monotonous, convergence patterns.

A second source of uncertainty is the network itself.
We can absorb this into the aforementioned distribution of predictions by training multiple FSPNs that have all been initialized randomly.
In this way, we obtain an even larger set of predictions that encapsulates both the sampling uncertainty and the network uncertainty.
Note that for the E2 moment both uncertainties are of similar size and the network uncertainty strongly varies for the different evaluation samples.
The network uncertainty might be improved through further optimization of the networks' topology and training process but is severely limited by the amount and quality of training data.
The sampling uncertainty demonstrates a sensitivity of the networks to the evaluation data, again emphasizing the importance of a thorough data selection.

\begin{figure}
    \centering
    \includegraphics[width=.8\columnwidth]{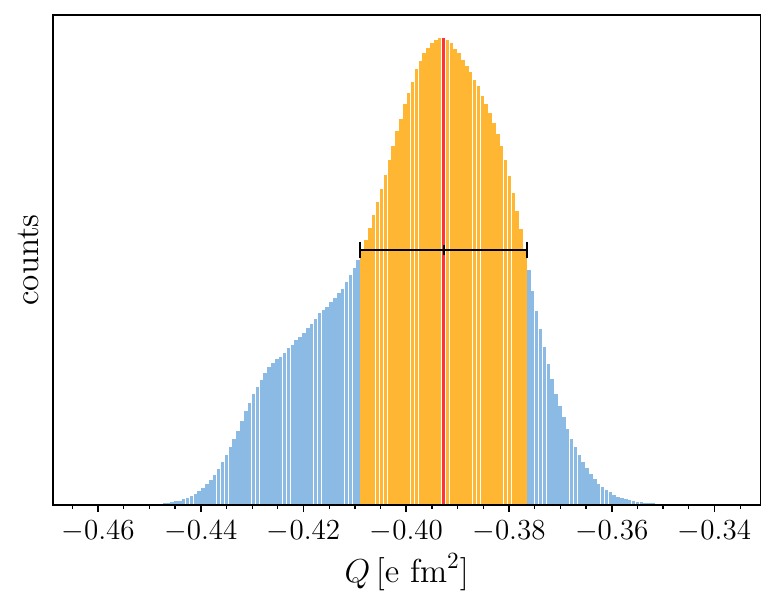}
    \caption{Distributions of predictions of the ground-state E2 moment in \elem{Li}{6} based on 1000 FSPNs evaluated with all possible samples at $\cNmax=12$ and a visualization of the statistical extraction of ensemble prediction and uncertainty.}
    \label{fig:statistical_evaluation}
\end{figure}
The remaining question is how to extract what we refer to as the ensemble prediction and uncertainty from the distribution.
We first depict the distribution as a histogram, where we employ the Freedman-Diaconis rule \cite{FrDi81} to estimate the bin size.
An example for such a histogram is shown in \cref{fig:statistical_evaluation} based on the predictions of the E2 moment of \elem{Li}{6} from 1000 FSPNs.
In order to avoid sensitivity to fluctuations, we first smooth out the histogram with a moving average over a width of ten bins.
Note that we typically obtain histograms with a few hundred bins, hence, ten bins is a rather small window and the overall shape of the histogram remains essentially unchanged.
We consider the ensemble prediction to be the most probable value, which is the highest bar in the histogram. 
Starting from this value we then add up the bars to the left and right of the ensemble prediction, always continuing with whichever is higher, until the accumulated number of counts exceeds \SI{68}{\percent} of the total number of predictions.
The resulting interval defines our uncertainty.
An illustration of this procedure is given by the orange interval in \cref{fig:statistical_evaluation}.
Hence, all uncertainties presented in this work are \SI{68}{\percent} confidence intervals.
The advantage of this extraction procedure over the Gaussian fit used in \cite{KnoWo23,WoKno24} is that it allows for asymmetric uncertainties.

\subsection{Results}
\begin{figure}[t]
    \centering
    \includegraphics[width=.99\columnwidth]{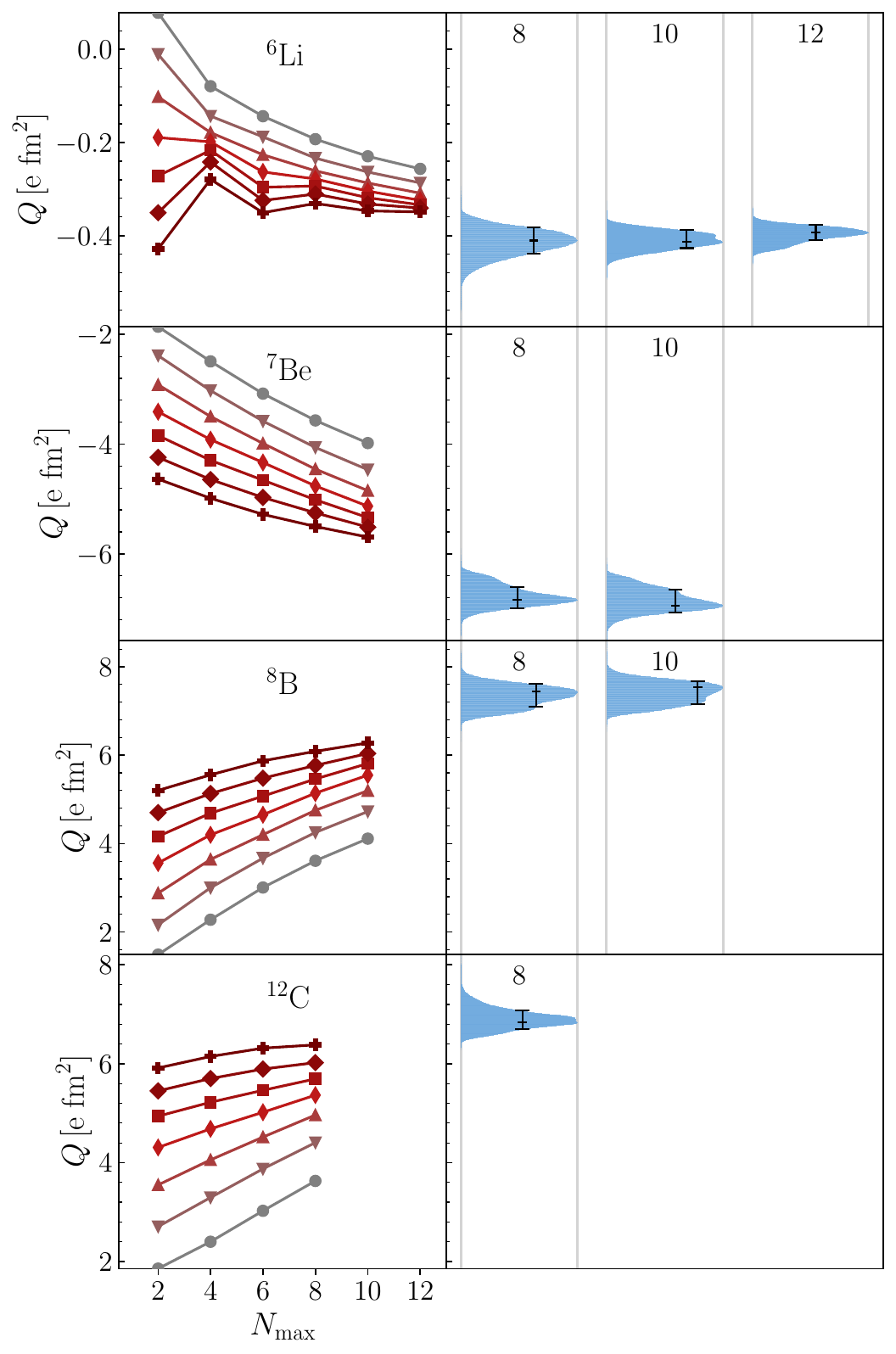}
    \caption{Input data and FSPN predictions from 1000 ANNs for the E2 momenta of the ground-states of \elem{Li}{6}, \elem{Be}{7}, and \elem{B}{8} and the first $2^+$ excited state of \elem{C}{12}.
    The left hand panels show NCSM data for $\aHO=1.2,1.3,1.4,1.5,1.6,1.7,$ and \SI{1.8}{\fm} (gray to red), while the right hand panels depict the predictions based this input data at different \cNmax.}
    \label{fig:FSPN_predictions}
\end{figure}
We can assess the quality of the FSPN approach for E2 moments based on the results for the ground-states of \elem{Li}{6}, \elem{Be}{7}, and \elem{B}{8} and the first $2^+$ excited state in \elem{C}{12}, which are shown in \cref{fig:FSPN_predictions}.
Here, the left-hand panels show the NCSM input data for the E2 moment over \Nmax, and with that the convergence pattern for the respective nucleus.
The right-hand panels depict the distributions of predictions from 1000 FSPNs for that input data at different \cNmax.
The selected nuclei feature different E2 moments and, more importantly, very different convergence patterns containing ascending, descending, and oscillating sequences.
Analogously to radii, the shape of the sequence strongly varies with \aHO.
Hence, the predictions depend on the \aHO present in the input data.
We, therefore, choose a window from $\aHO=\SIrange{1.2}{1.8}{\fm}$ corresponding to $\hw=\SIrange{12.8}{28.8}{\fm}$, which is within the \hw range of the training data.
All evaluation data are computed with the non-local chiral NN + 3N interactions from \cite{EnMa17,HuVo20} at N$^3$LO with cutoff $\Lambda = \SI{500}{\MeV}$, which we refer to as EMN[500], and SRG evolved to $\alpha=\SI{0.08}{\fm\tothe{4}}$.

Starting with \elem{Li}{6}, we find that the overall convergence pattern is similar to the \elem{H}{2} training data except for the direction of convergence, since the E2 moment of \elem{Li}{6} is negative.
As this is not represented in the training data, the sign of the evaluation samples, and with that the direction of convergence, is flipped before they are passed through the FSPNs.
In \elem{Li}{6} we find very consistent predictions across all \cNmax. 
Though, they appear somewhat large in magnitude compared to the most converged sequence at $\aHO=\SI{1.8}{\fm}$.
A similar picture emerges for \elem{Be}{7}, which exhibits a monotonously decreasing convergence pattern. 
Finally, for \elem{B}{8} and \elem{C}{12} we have a monotonously increasing convergence pattern that is akin to the training data at large \Nmax.
Once more, find very consistent predictions, however, in the case of \elem{B}{8} they feature a deformed, almost bimodal distribution, which generally points at a mismatch of input data and training data.

Overall, the results obtained with the FSPN approach are surprisingly good, considering the very limited training data, and demonstrate the predictive power and universal applicability of this method including excited states.

However, the FSPN approach is not applicable to other EM observables due to the complete lack of suitable training data from few-body systems.
This limitation will remain until high-quality converged training data can be obtained for light p-shell nuclei, which is out of reach for now.
Hence, we need to find an alternative approach in order to predict the full-space values of these observables. 

\section{Observable Transcoder Networks}\label{sec:OTN}

In order to overcome the limitations of the FSPNs, we propose OTNs as a new machine learning approach that leverages the correlations between observables.
Instead of characterizing the model space with the NCSM specific parameters \Nmax and \hw, we substitute the energy for \Nmax and replace \hw with the radius.
As both observables depend on \Nmax and \hw this resembles a reparametrization, where the radius mostly carries information on the length scale of the HO basis, due to its sensitivity to \aHO, and the energy yields information on the model-space size due to monotonicity.
Thus, we essentially aim for a mapping $(E,R) \rightarrow O$ that converts energies and radii into the observable of interest.

Here, we use the energy and the point-proton radius to train the OTNs to predict the E2 moment from these inputs.
Hence, we aim to replace the model-space extrapolation with an observable conversion by exploiting the correlations between the different observables.
In order to obtain a prediction for the full-space E2 moment we need to supply the OTN with the full-space energy and radius.
In Ref. \cite{WoKno24} we have shown that FSPNs can provide high-quality full-space predictions for these observables.

\begin{figure}
    \centering
    \includegraphics[width=\columnwidth]{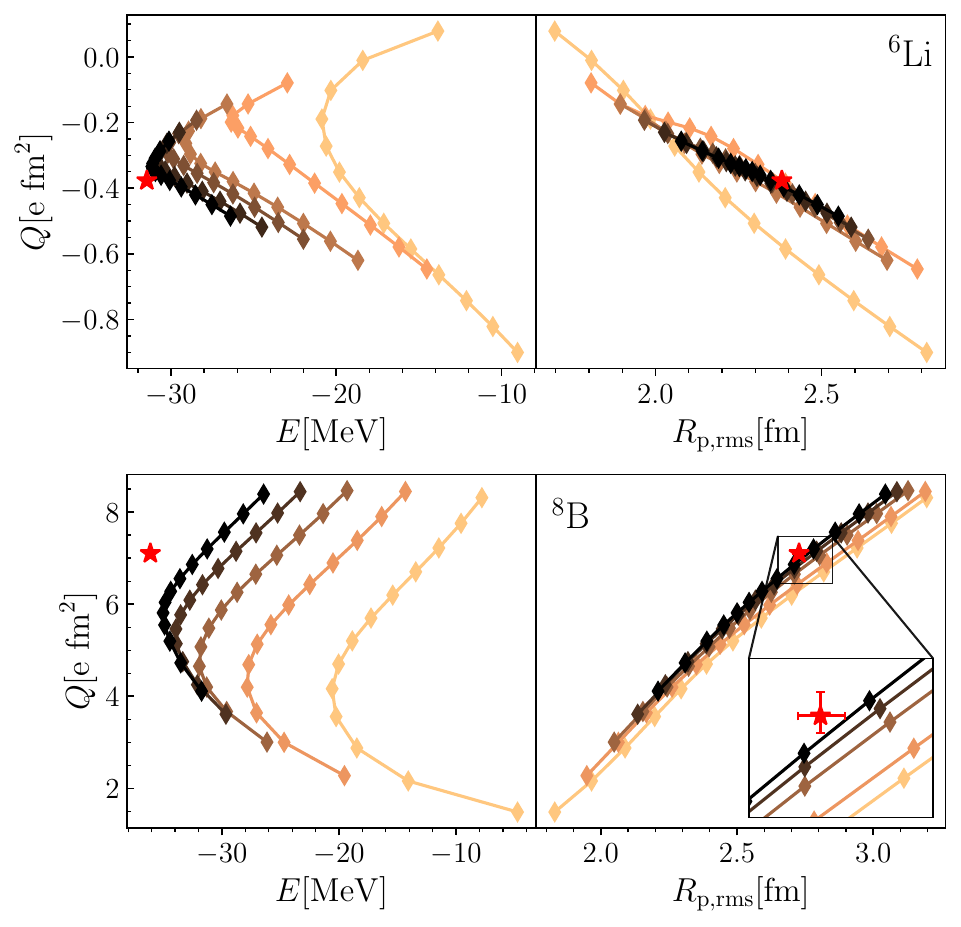}
    \caption{Correlations between the ground-state E2 moment and the ground-state energy (left) and the ground-state point-proton radius (right) in \elem{Li}{6} and \elem{B}{8} for EMN[500] from $\Nmax=2$ (yellow) to $\Nmax=12/10$ (black). The curves for each \Nmax are spanned by an \aHO range from \SIrange{1.2}{2.4}{\fm}. The red stars illustrate the FSPN predictions for the energies or radii, and the OTN predictions for the E2 moments, with their respective error bars.}
    \label{fig:correlations}
\end{figure}

Before we discuss the OTNs in detail, let us first study the correlations between the E2 moment and energy or point-proton radius for a specific nuclear state.
An example of these correlations for the ground states of \elem{Li}{6} and \elem{B}{8} at different \Nmax is given in \cref{fig:correlations}.
The right-hand panels show the well-known correlation between the E2 moment and the radius, which both strongly depend on \aHO.
This correlation becomes obvious when comparing the operators for E2 moment and point-proton radius. 
This similarity has already been exploited in other work regarding E2 moments and transitions in the NCSM \cite{CaFa22,CaMa24}.
Note that the correlations are already present in small model spaces and converge rapidly.

\begin{figure*}
    \centering
    \includegraphics[width=\textwidth]{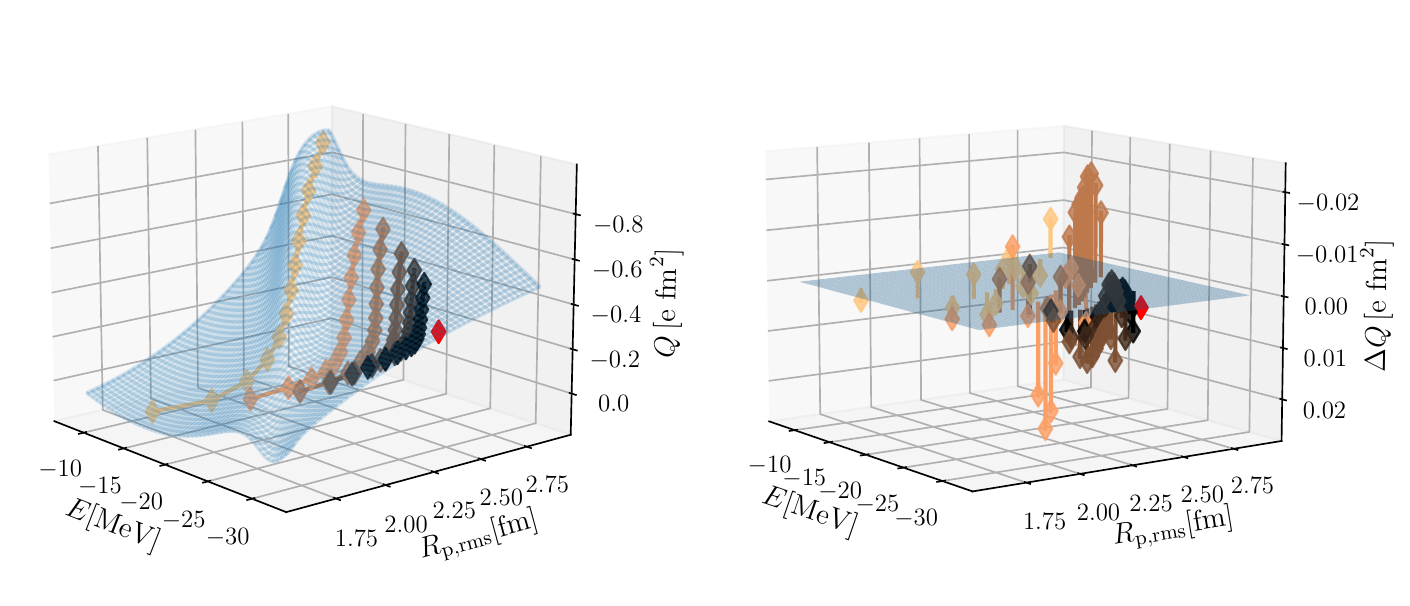}
    \caption{Left: The OTN training data for the ground state of \elem{Li}{6} (same as in \cref{fig:correlations}) along with a grid of predictions from a trained OTN that emulates the surface spanned by the training data. Right: Deviations between the NCSM data and the OTN predictions as a measure of accuracy in the description of the training data. The red marker represents the OTN prediction for the E2 moment based on FSPN predictions for energy and point-proton radius in order to illustrate its proximity to the training data.}
    \label{fig:3Dsurface}
\end{figure*}

The left-hand panels of \cref{fig:correlations} show that there are no apparent correlations between E2 moment and energy.
Hence, we expect the OTNs to predominantly learn the radius correlation.
Nevertheless, we keep the energy in the input layer as a measure of convergence and with the extension to other EM observables in mind.

\subsection{Network Design}

Modelling the correlations is less complex than the pattern matching approach of the FSPNs. 
Hence, a smaller topology is sufficient for the OTNs.
A schematic depiction is provided in \cref{fig:topology}.
The size of input and output layers are naturally given by the desired mapping.
Thus, the input layer contains two input nodes, one for energy and one for radius, and a single output node that yields the prediction of the E2 moment.
The OTNs further consist of two hidden layers with 8 nodes each, resulting in 107 trainable parameters.
For the activation function we have found a hyperbolic tangent to yield best results, while we stick with an MSE loss function and the AdamW backpropagation algorithm. 

In order to obtain training data for the OTNs a method to probe the three-dimensional space spanned by energy, radius, and E2 moment is required.
Moreover, the correlations are unique to the combination of nucleus, Hamiltonian and state under investigation. 
This has two major implications: 
First, the OTNs do not share the universality of the FSPNs and need to be retrained in every application.
Second, a sufficient amount of training data needs to be accessible for a given nucleus and state.
Under these considerations the NCSM is the ideal method to obtain high-quality training data from ab initio theory since all observables are directly accessible, the amount of training data can be increased by a variation of \aHO, and as demonstrated in \cref{fig:correlations} the dominant correlations already emerge in small model spaces.

The left-hand side of \cref{fig:3Dsurface} depicts the NCSM training data for the ground state of \elem{Li}{6} for $\aHO = 1.2,1.3,\ldots,\SI{2.4}{\fm}$ and $\Nmax = 2$ to $12$. 
Note that $\Nmax=0$ is discarded due to limited physical relevance. 
The data spans a smooth surface that is non-degenerate w.r.t.\ the E2 moment and can, therefore, easily be modeled with an ANN.
This is further simplified by normalizing the data to the interval $[0,1]$ for all three observables.
We train our OTN on this data for a total of 10\,000 epochs with a batch size of 4 and an initial learning rate of 0.001 which is successively reduced during the training by a factor of 0.8 whenever the loss plateaus, i.e.\ it does not decrease by at least \SI{0.1}{\percent} for 80 epochs.
We consider the training successful and the OTN valid if the MSE loss becomes smaller than \SI{0.0001}{}.
The whole training process takes less than a minute and multiple networks can be trained simultaneously.

We can assess the correlations learned by the OTN by probing the observable space and illustrating the surface predicted by the network as a blue grid, which is also shown in \cref{fig:3Dsurface}.
In addition, the right-hand side shows the deviation between training data and OTN predictions.
We find that the network is capable to emulate the training data to good precision especially in larger \Nmax.
Once trained, the OTN can be applied to unseen data.

We emphasize that, compared to the FSPN approach, the output of the OTN is not a prediction for the full-space E2 moment but the E2 moment corresponding to the model space in which the energy and radius results have been obtained. 
Consequently, the OTN can provide a prediction for the full-space E2 moment when the input energy and radius are results or predictions for the values in the full Hilbert space.
While these results for energy and radius can be obtained by other means, we will use FSPN predictions as input to the OTNs, as they have proven to provide reliable estimates for the full-space energy and radius.
Thus, we obtain full-space predictions for the E2 moment via a two-step process that combines both ANN architectures.
In a first step, we perform NCSM calculations for energy, radius, and E2 moment and extrapolate energy and radius with the FSPNs from \cite{WoKno24}. 
Subsequently, we train OTNs on the NCSM data and use the FSPN predictions for energy and radius as an input to these OTNs, which then yield predictions for the E2 moment.

The red symbols in \cref{fig:correlations,fig:3Dsurface} show examples for such predictions obtained from the depicted data.
It demonstrates the proximity of the prediction to the training data.
This is important to note since extrapolation problems are generally difficult for ANNs, while they excel on interpolation tasks.
Due to the wide \aHO grid, we find that the OTN effectively performs an interpolation of the E2 moment w.r.t.\ the radius and the small extrapolation w.r.t.\ energy is of minor importance since E2 moment and energy are not as correlated.
Nevertheless, the inset in \cref{fig:correlations} demonstrates the relevance of the energy as an input to the OTN as it allows the network to account for the remaining \Nmax dependence of the correlation curve.

Overall, we find that the OTNs yield an accurate model of the correlations between the different observables and allow for predictions of E2 moments.

\subsection{Uncertainty Propagation}

\begin{figure}
    \centering
    \includegraphics[width=.8\columnwidth]{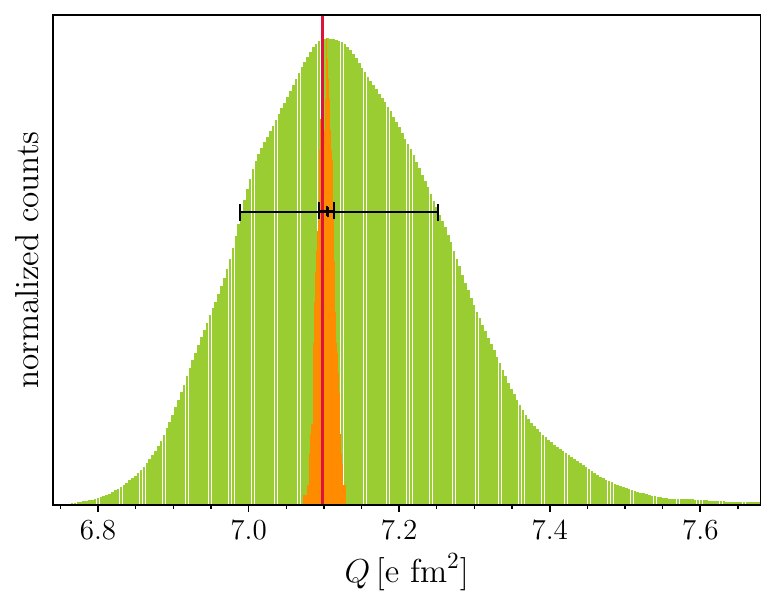}
    \caption{Distribution of predictions of the ground-state E2 moment of \elem{B}{8} based on 100 OTNs evaluated with the FSPN ensemble predictions for energy and point-proton radius (orange) and evaluated with 10,000 samples randomly drawn from the distributions of FSPN predictions for the two observables (green). Error bars indicate the respective ensemble predictions and uncertainties. The vertical red line illustrates the prediction of a single OTN for comparison.}
    \label{fig:uncertainty_propagation}
\end{figure}

In order to obtain meaningful predictions for the E2 moment we need to account for different sources of uncertainty, in particular, the uncertainty of the energy and radius inputs as well as the network uncertainty of the OTN.
In analogy to the FSPN network uncertainty, we can account for the latter by training a total of 100 OTNs with different random initializations of the network parameters. 
\Cref{fig:uncertainty_propagation} illustrates how the prediction from a single OTN shown in red gets replaced by the orange histogram that now includes predictions from 100 OTNs.
The small size of the extracted OTN uncertainty is an indication that the networks can successfully account for the minor extrapolation in the energy direction.
We further need to consider the uncertainties for energy and radius from which the E2 moment is predicted. 
These uncertainties are encoded in the histograms of predictions obtained from the FSPN approach.
By sampling these distributions a total of 10\,000 times, each yielding a pair of randomly selected energy and radius, and evaluating the 100 OTNs with all of these samples the uncertainties can be propagated to the E2 moment.
The resulting histogram shown in green now includes all sources of uncertainty on the many-body level.
The FSPN predictions of the radius are by far the dominant source of uncertainties for the E2 moment, while the OTN uncertainty is negligibly small in comparison.
This is consistent with the strong correlation between the two observables and, therefore, the precision of the OTN predictions is ultimately limited by the precision of the FSPN predictions for the radius.

\subsection{Results}

With the discussed procedure for prediction and uncertainty quantification for E2 moments, we can investigate the quality of the OTN predictions for selected p-shell nuclei.
The NCSM calculations for these investigations have been obtained with the  EMN[500] interaction and the Hamiltonian further is SRG evolved to $\alpha = \SI{0.08}{\fm\tothe{4}}$.

\begin{figure}
    \centering
    \includegraphics[width=.99\columnwidth]{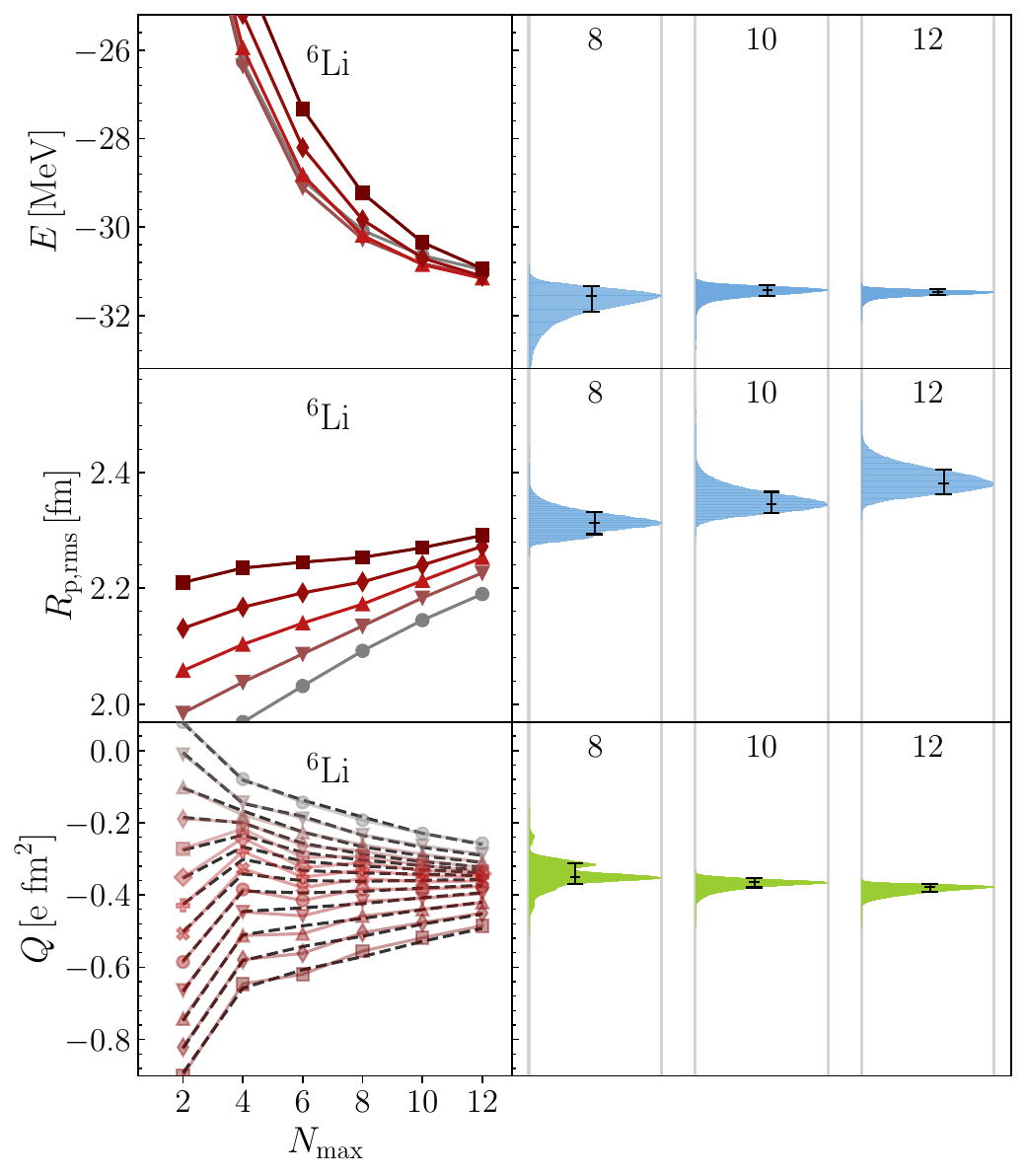}
    \caption{Input data and FSPN predictions (blue) for the ground-state energy (top panels) and point-proton rms radius (center panels) of \elem{Li}{6} with the EMN[500] interaction. The bottom panels show a comparison of the NCSM results for the E2 moment and the OTN predictions (green) obtained from the FSPN predictions at the respective \cNmax. The dashed black lines in the bottom left panel additionally show the OTNs performance on the depicted NCSM training data.}
    \label{fig:OTN_predictions_Li6}
\end{figure}

We start with the ground state of \elem{Li}{6} for which both, the FSPN predictions for energy and point-proton radius along with the OTN predictions for the E2 moment are depicted in \cref{fig:OTN_predictions_Li6}.
We find that the energy predictions are very stable and consistent across the different \cNmax.
The predictions of the point-proton radius, however, show some trend to larger values with increasing \cNmax due to the anomalous change in the convergence pattern starting from $\Nmax=10$, best visible in the highest-lying sequence which appears to be converging before changing to an almost linear increase.
This effect is particularly pronounced in \elem{Li}{6} because of its cluster structure, which allows us to investigate if and how such deficiencies in the radius extrapolation translate to E2 moments.

Next, we assess the performance of the OTNs on the training data shown by the dashed black lines in the bottom left panel.
We find that the description of the data is very accurate across all \aHO and \Nmax. 
The OTN slightly smooths out the alternating behavior of some sequences in $\Nmax=4,6$.
This indicates that the networks capture the dominant correlations without overfitting all artifacts in the data.
With this additional quality assessment, we can turn to extractions of the E2 moments obtained from the FSPN predictions at different \cNmax.
Note that the \cNmax truncation affects both the FSPN predictions of energy and radius and the OTN training data.
We find that the predictions are very precise compared to the spread of the NCSM data given in the bottom left panel and feature slightly smaller uncertainties than the direct FSPN predictions in \cref{fig:FSPN_predictions}.
This is a result of the precise radius predictions along with the strong and accurately modeled correlations.
Upon closer inspection we find that the linear trend in the radius predictions does, in fact, translate to the predictions of the E2 moments but on a reduced scale. 

From this discussion of the particularly challenging \elem{Li}{6} we can conclude that, besides a nontrivial convergence pattern and a significant trend in the radius predictions, the OTNs provide precise and robust predictions for the E2 moment.
Hence, we expect them to perform as good or even better on other nuclei that show a more regular behavior.

\begin{figure}
    \centering
    \includegraphics[width=.99\columnwidth]{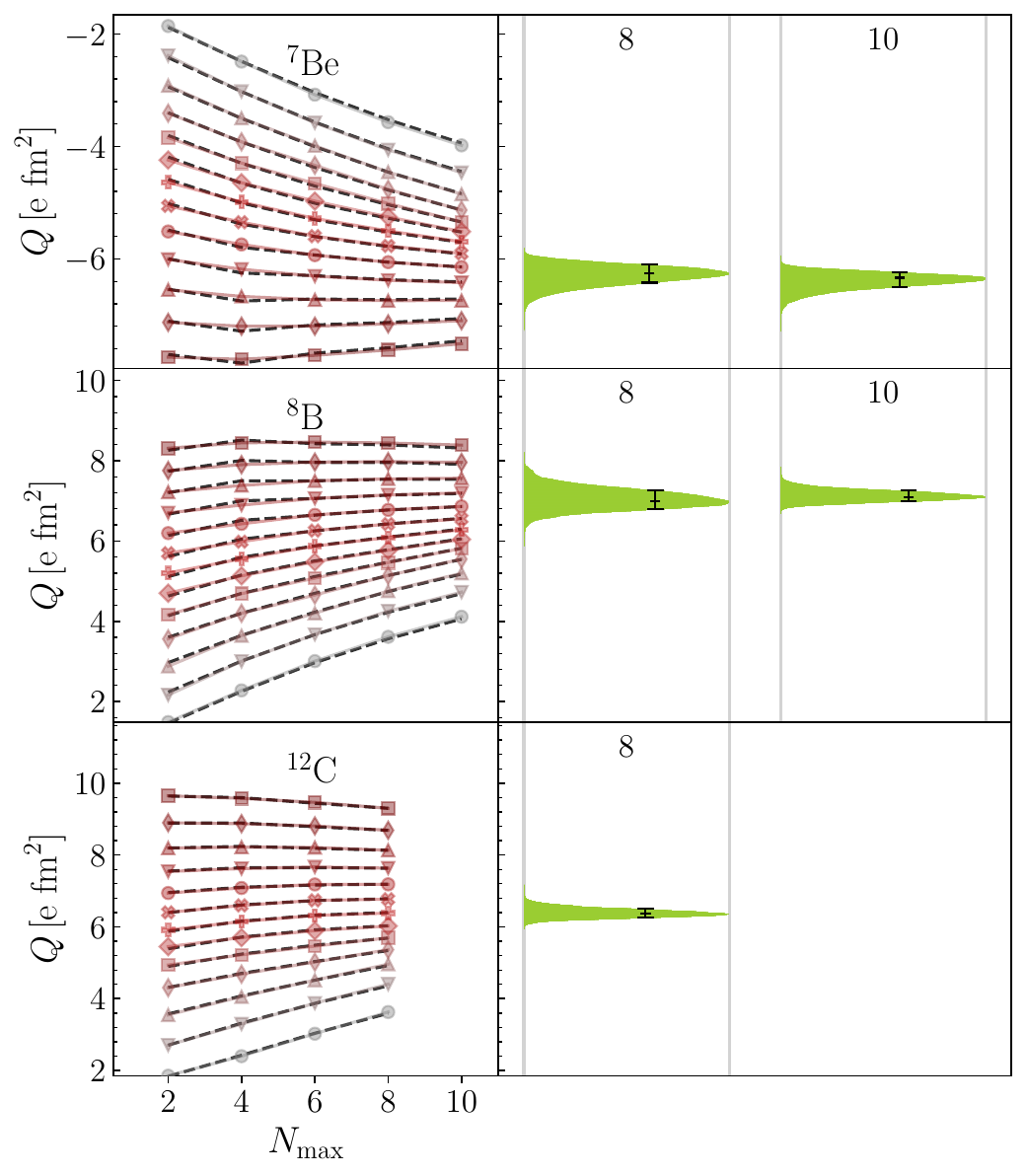}
    \caption{Comparison of the NCSM training data for EMN[500] along with OTN predictions as black dashed lines (left-hand panels) with the OTN predictions from FSPN predictions of energy and point-proton radius (right-hand panels) for the ground-state of \elem{Be}{7} and \elem{B}{8} and for the first $2^+$ excited state in \elem{C}{12}.}
    \label{fig:OTN_predictions}
\end{figure}

The ground states of \elem{Be}{7} and \elem{B}{8} and the first $2^+$ excited state in \elem{C}{12} are examples of nuclei with a more regular convergence of radius and E2 moment. 
The respective OTN predictions for the E2 moments are given in \cref{fig:OTN_predictions}.
Again, the left-hand panels show the NCSM training data along with the OTN predictions based on the actual NCSM energy and radius for the respective model space. The OTN predictions based on the FSPN results for the converged
energy and radius at different \cNmax are given in the right-hand panels.
We first notice that the OTNs describe the training data very accurately.
Further, we find precise predictions in all nuclei that are consistent across different \cNmax, which is a result of consistent predictions for the respective radii.
Once more, we highlight the precision of the predictions compared to the wide spread of the converging sequences in the NCSM data. 

In order to assess the accuracy of the OTN approach we would ideally compare the predictions to the exact results.
Unfortunately, a fully converged E2 moment is only accessible in \elem{H}{2}, which exhibits an atypical staggering in the convergence pattern as demonstrated in \cref{fig:training_data}.
This translates to a complicated structure in the correlation surface that is to be modeled by the OTNs.
Thus, an application to \elem{H}{2} 
is not representative for the p-shell nuclei discussed here.
However, we can assess the accuracy of the OTN approach indirectly.
On one hand, the FSPNs account for the model-space extrapolation and the accuracy of the FSPN approach is discussed in detail in Refs.~\cite{WoKno24,KnoLo25}.
On the other hand, as already discussed, we find that the OTNs accurately model the smooth correlation surfaces for p-shell nuclei and the spread of predictions from different OTNs is very small compared to the radius uncertainties.
All these arguments speak to the accuracy of the OTN approach.

\subsection{Interaction Uncertainties}

\begin{figure}
    \centering
    \includegraphics[width=.99\columnwidth]{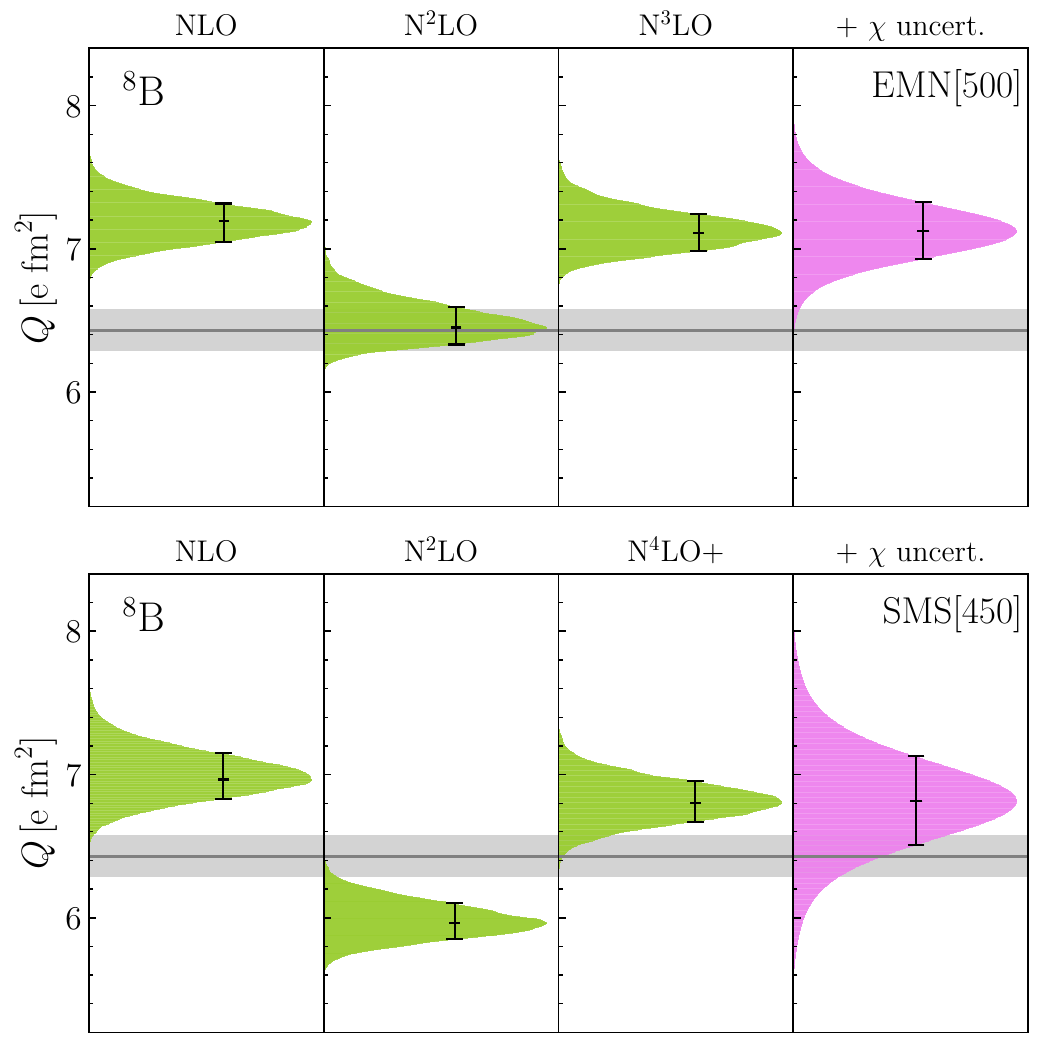}
    \caption{OTN predictions for \elem{B}{8} at different chiral orders (green) and at the highest order with included chiral uncertainties (violet) for the EMN[500] (top panel) and SMS[450] (bottom panel) interactions. The experimental result from \cite{Stone16} is given in grey for comparison.}
    \label{fig:B8_full_uncert}
\end{figure}

For a meaningful comparison with experiment, we also need to address the interaction uncertainties that arise from the truncation of the chiral expansion.
For this we employ the Bayesian point-wise model for interaction uncertainties in chiral EFT developed by the BUQEYE collaboration \cite{MeFu19}.
Hence, we need to perform NCSM calculations for energy, radius, and E2 moment as well as the ANN evaluates at the different chiral orders.
The ensemble predictions at the different orders are used as input to the BUQEYE model that provides a student-t distribution for the highest order that encapsulates the interaction uncertainty.
In order to combine many-body and interaction uncertainties, we compute the convolution of the student-t distribution with the histogram of OTN predictions at the highest order accessible.
This yields a final distribution around a similar nominal value but with a wider spread that now resembles a full uncertainty (see \cite{WoGe25} for details).

\Cref{fig:B8_full_uncert} illustrates this procedure and shows the results for \elem{B}{8} for the EMN[500] interaction at NLO, N$^2$LO, and N$^3$LO and additionally for the semi-local NN+3N interaction from \cite{ReKr18,LENPIC21,Mar22} with cutoff $\Lambda = \SI{450}{\MeV}$ labeled as SMS[450] at NLO, N$^2$LO, and N$^4$LO+.
All Hamiltonians have been SRG evolved to $\alpha=\SI{0.08}{\fm\tothe{4}}$.
We find that both interactions share the same systematics, in particular a shift to smaller values in N$^2$LO that is remedied in the next order.
Note that the chiral uncertainties are larger for the SMS[450] interaction as 3N forces are only included up to N$^2$LO and we, thus, choose to treat  N$^4$LO+ as another third order result in the BUQEYE model.
The final predictions for EMN[500] overestimate the experimental value, the SMS[450] also tend to larger E2 moments, but are in agreement with experiment within uncertainties.

\begin{figure}
    \centering
    \includegraphics[width=.99\columnwidth]{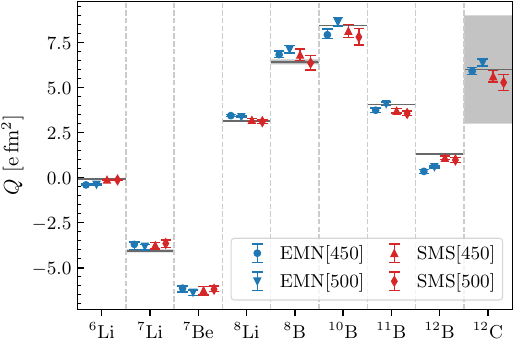}
    \caption{OTN predictions for the E2 moment with combined many-body and chiral interaction uncertainties for various p-shell nuclei and the EMN and SMS interactions at N$^3$LO and N$^4$LO+, respectively, for cutoffs $\Lambda=450$ and $500$~\si{\MeV}. Experimental data from \cite{Stone16} is shown for comparison.}
    \label{fig:summary}
\end{figure}

We now extend our investigations to a wide range of p-shell nuclei, i.e., \elem{Li}{6,7,8}, \elem{Be}{7}, \elem{B}{8,10,11,12}, and \elem{C}{12}.
The final results with complete uncertainty quantification for four different interactions, in particular the EMN and SMS families with two cutoffs $\Lambda=450$ and \SI{500}{\MeV} each, are depicted in \cref{fig:summary} along with E2 moments from experiment.
All results and experimental values are further given in \cref{tab:summary}.
For all nuclei except \elem{Li}{6} and \elem{B}{12} the results are in good agreement with experiment within uncertainties for at least one of the interactions.
Considering that our uncertainties are constructed to resemble a 1-$\sigma$ interval, the results for \elem{Li}{6} and \elem{B}{12} fall within a 2-$\sigma$ interval and are, thus, also within the range of the respective distributions of predictions.
Overall the SMS interactions yield slightly more accurate results. 
However, we find some variations with cutoff in both families that corresponds to the change in the respective radii.
This effect is more pronounced for the EMN interactions, while the SMS interactions are in agreement within uncertainties.
Due to the small E2 moment in \elem{Li}{6} this results in a large relative deviation, still, all predictions remain in qualitative agreement with experiment.

\begin{table}[t]
    \centering
    \begin{ruledtabular}
    \begin{tabular}{c c c r r c}
         {{\multirow{2}{*}{Nucl.}}} & {{\multirow{2}{*}{\Nmax}}} & {{\multirow{2}{*}{Int.}}} & \multicolumn{2}{c}{$Q$~[\si{\quadrupole}]} & {{\multirow{2}{*}{Exp.}}} \\
         \cline{4-5} \\[-1em]
         & & & \multicolumn{2}{c}{$\Lambda=450$~\si{\MeV}\quad\;$500$~\si{\MeV}}\\
         \hline\\[-1em]
         & & & &\\[-.5em]
         {{\multirow{2}{*}{\elem{Li}{6}}}} & {{\multirow{2}{*}{12}}} & {{EMN}} & $-0.402_{-0.046}^{+0.044}$ & $-0.384_{-0.046}^{+0.044}$  & \multirow{2}{*}{\tablenum[table-format=2.6]{-0.0806(6)}} \\[.5em]
         & & {{SMS}} & $-0.121_{-0.031}^{+0.023}$ & $-0.124_{-0.022}^{+0.018}$ & \\[.5em]
         \hline\\[-1em]
         & & & &\\[-.5em]
         {{\multirow{2}{*}{\elem{Li}{7}}}} & {{\multirow{2}{*}{10}}} & {{EMN}} & $-3.700_{-0.299}^{+0.139}$ & $-3.816_{-0.191}^{+0.122}$  & \multirow{2}{*}{\tablenum[table-format=2.6]{-4.06(8)}} \\[.5em]
         & & {{SMS}} & $-3.756_{-0.198}^{+0.157}$ & $-3.649_{-0.215}^{+0.188}$ & \\[.5em]
         \hline\\[-1em]
         & & & & \\[-.5em]
         {{\multirow{2}{*}{\elem{Be}{7}}}} & {{\multirow{2}{*}{10}}} & {{EMN}} &$-6.130_{-0.205}^{+0.129}$ & $-6.369_{-0.163}^{+0.145}$ &  {{\multirow{2}{*}{n.a.}}} \\[.5em]
         & & {{SMS}} & $-6.262_{-0.253}^{+0.227}$ & $-6.190_{-0.139}^{+0.173}$ & \\[.5em]
         \hline\\[-1em]
         & & & &\\[-.5em]
         {{\multirow{2}{*}{\elem{Li}{8}}}} & {{\multirow{2}{*}{10}}} & {{EMN}} & $3.443_{-0.053}^{+0.060}$ & $3.374_{-0.034}^{+0.046}$ & \multirow{2}{*}{\tablenum[table-format=2.6]{3.14(2)}} \\[.5em]
         & & {{SMS}} & $3.193_{-0.081}^{+0.082}$ & $3.107_{-0.102}^{+0.105}$ & \\[.5em]
         \hline\\[-1em]
         & & & & \\[-.5em]
         {{\multirow{2}{*}{\elem{B}{8}}}} & {{\multirow{2}{*}{10}}} & {{EMN}} & $6.827_{-0.167}^{+0.192}$ & $7.124_{-0.196}^{+0.203}$ & \multirow{2}{*}{\tablenum[table-format=2.6]{6.43(14)}} \\[.5em]
         & & {{SMS}} & $6.814_{-0.306}^{+0.314}$ & $6.367_{-0.393}^{+0.400}$ & \\[.5em]
         \hline\\[-1em]
         & & & &\\[-.5em]
         {{\multirow{2}{*}{\elem{B}{10}}}} & {{\multirow{2}{*}{8}}} & {{EMN}} & $7.930_{-0.216}^{+0.315}$ & $8.637_{-0.211}^{+0.246}$ & \multirow{2}{*}{\tablenum[table-format=2.6]{8.45(2)}} \\[.5em]
         & & {{SMS}} & $8.136_{-0.348}^{+0.353}$ & $7.805_{-0.451}^{+0.467}$ & \\[.5em]
         \hline\\[-1em]
         & & & &\\[-.5em]
         {{\multirow{2}{*}{\elem{B}{11}}}} & {{\multirow{2}{*}{8}}} & {{EMN}} & $3.742_{-0.113}^{+0.122}$ & $4.091_{-0.089}^{+0.105}$ & \multirow{2}{*}{\tablenum[table-format=2.6]{4.059(10)}} \\[.5em]
         & & {{SMS}} & $3.704_{-0.122}^{+0.127}$ & $3.564_{-0.127}^{+0.130}$ & \\[.5em]
         \hline\\[-1em]
         & & & &\\[-.5em]
         {{\multirow{2}{*}{\elem{B}{12}}}} & {{\multirow{2}{*}{8}}} & {{EMN}} & $0.350_{-0.105}^{+0.089}$ & $0.590_{-0.053}^{+0.063}$ & \multirow{2}{*}{\tablenum[table-format=2.6]{1.32(3)}} \\[.5em]
         & & {{SMS}} & $1.095_{-0.099}^{+0.101}$ & $0.981_{-0.133}^{+0.133}$ & \\[.5em]
         \hline\\[-1em]
         & & & & \\[-.5em]
         {{\multirow{2}{*}{\elem{C}{12} ($2^+$)}}} & {{\multirow{2}{*}{8}}} & {{EMN}} & $5.910_{-0.188}^{+0.215}$ & $6.380_{-0.177}^{+0.194}$ & \multirow{2}{*}{\tablenum[table-format=2.6]{6(3)}} \\[.5em]
         & & {{SMS}} & $5.615_{-0.310}^{+0.319}$ & $5.281_{-0.442}^{+0.445}$ & \\[.5em]
    \end{tabular}
    \end{ruledtabular}
    \caption{OTN predictions for the E2 moment with included chiral uncertainties for various p-shell nuclei and the EMN and SMS interactions at N$^3$LO and N$^4$LO+, respectively, for cutoffs $\Lambda=450$ and $500$~\si{\MeV}. The second column indicates the highest \Nmax available for the respective nucleus. Experimental data from \cite{Stone16} is given for comparison.}
    \label{tab:summary}
\end{table}

\subsection{Towards Cross-Method Applications}

The OTNs offer an interesting perspective for extensions to heavier nuclei.
Being designed to capture the correlations between observables, they do not rely on method-specific parameters and are, therefore, agnostic to the source of their input.
Hence, the energy and radius used to evaluate the OTNs can, in principle, be obtained through any many-body method as long as the same Hamiltonian is employed.
In particular, this allows us to train OTNs based on NCSM data and evaluate them with energies and radii obtained from many-body methods that can access heavier nuclei.
It further eliminates the need to directly compute EM observables in these models.

\begin{figure}
    \centering
    \includegraphics[width=.99\columnwidth]{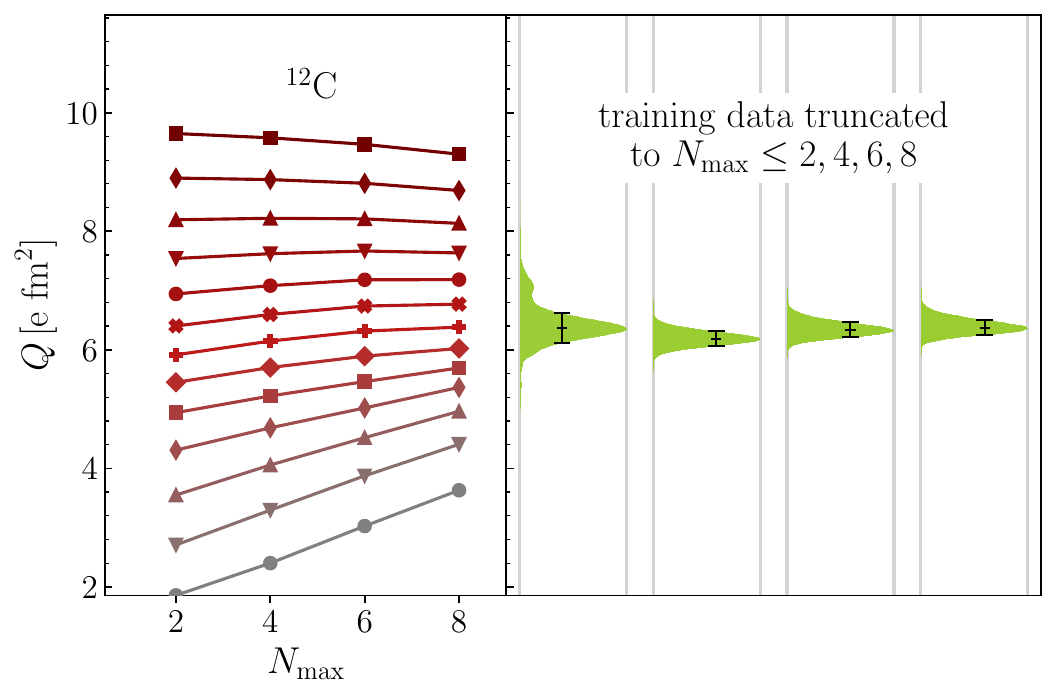}
    \caption{NCSM data (left-hand panel) along with OTN predictions (right-hand panel) for the first $2^+$ excited state in \elem{C}{12} based on OTNs trained with different sets of training data truncated to $\Nmax=2,4,6,$ and $8$, where the latter corresponds to the full training data available.}
    \label{fig:OTN_predictions_C12_truncated}
\end{figure}

Eventually, the availability of training data from the NCSM becomes the bottleneck of this kind of cross-method applications as only very small \Nmax can be reached for, e.g., sd-shell nuclei.
Thus, we investigate how much training data is actually required in order to reliably construct a set of OTNs.
We simulate this shortage of data by artificially truncating the training set for the OTNs to smaller \Nmax.
\Cref{fig:OTN_predictions_C12_truncated} shows predictions for the E2 moment of the first excited $2^+$ state in \elem{C}{12} obtained with different OTNs that have been trained on NCSM data truncated to $\Nmax\leq2,4,6,$ and $8$, respectively.
We find that, besides the extremely limited $\Nmax=2$ case, the predictions are remarkably consistent and very little training data is sufficient to capture the correlations.
This aligns with the rapid convergence of the E2 moment and radius correlation found in \cref{fig:correlations}. 
Hence, reliable OTNs can likely be constructed up into the mid sd-shell.

\section{Conclusions}
In this work, we have extended our previous FSPN approach for the extrapolation of NCSM calculations to E2 moments and developed a new OTN framework that predicts E2 moments from energies and radii based on correlations learned from ab initio calculations.
While both approaches yield compatible results for the selected p-shell nuclei, the FSPNs are more prone to unwanted biases due to the lack of training data and this limitation will ultimately prevent the extension to other EM observables.
The OTNs, however, exploit the strong correlation between the E2 moment and the radius resulting in robust and precise predictions of E2 moments based on FSPN predictions for energies and radii.
These correlations can be modeled to such a degree that the precision of the OTN approach is only limited by the precision of the radius extrapolation.

In addition to a many-body uncertainty, we have demonstrated how interaction uncertainties from the truncation of the chiral expansion can be incorporated in the many-body predictions, thus, allowing for a complete statistical uncertainty quantification.
This enables a precision study of E2 moments in p-shell nuclei that yields important insight into details of the nuclear Hamiltonian.

Finally, the OTN approach is directly applicable to other EM observables and can naturally be extended to other many-body methods. 
It even shows high potential for cross-method applications, which can increase the reach towards sd-shell nuclei and eliminate the need to compute EM observables in medium-mass methods.
Work along these lines is in progress.

\vspace{2em}

\begin{acknowledgments}

This work is supported by the Deutsche\linebreak Forschungsgemeinschaft (DFG, German Research Foundation) through the DFG Sonderforschungsbereich SFB 1245 (Project ID 279384907), 
the BMBF through Verbundprojekt 05P2024 (ErUM-FSP T07, Contract No. 05P24RDB),
and by the U.S. Department of Energy, Office of Science, under Award Number DE-SC0023495 (SciDAC5/NUCLEI).
The NCSM calculations for the results presented here were performed with the code MFDn \cite{ccpe-10-2013-Aktulga,SHAO20181}
on Theta and Aurora at the Argonne Leadership Facility (ALCF), a DOE Office of Science User Facility supported by the Office of Science of the U.S. Department of Energy under Contract DE-AC02-06CH11357, under the Innovative and Novel Computational Impact on Theory and Experiment (INCITE) program;
and on Perlmutter at the National Energy Research Scientific Computing Center (NERSC), a DOE Office of Science User Facility supported by the Office of Science of the U.S. Department of Energy under Contract No. DE-AC02-05CH11231, using NERSC award ERCAP0028672.
Numerical calculations for the training data have been performed on the LICHTENBERG II cluster at the computing center of the TU Darmstadt.
\end{acknowledgments}

\bibliography{bib_nucl}

\end{document}